\documentclass[aps,reprint,prb,superscriptaddress]{revtex4-1}

\usepackage{bm}
\usepackage{amssymb}
\usepackage{amsmath}
\usepackage{color}
\usepackage{multirow}
\usepackage{tabularx}
\usepackage{graphicx,epsfig}

\begin{document}

\title{Angle-Adjustable Density Field Formulation for Modeling Crystalline Microstructures}
\author{Zi-Le Wang}
\affiliation{College of Chemistry and Molecular Engineering, Peking University,
Beijing 100871, China}
\author{Zhirong Liu}
\email{liuzhirong@pku.edu.cn}
\affiliation{College of Chemistry and Molecular Engineering, Peking University,
Beijing 100871, China}
\author{Zhi-Feng Huang}
\email{huang@wayne.edu}
\affiliation{Department of Physics and Astronomy, Wayne State University,
Detroit, Michigan 48201, USA}

\date{\today}

\begin{abstract}
  A continuum density-field formulation with particle-scale resolution is
  constructed to simultaneously incorporate the orientation dependence of
  interparticle interactions and the rotational invariance of the system,
  a fundamental but challenging issue in modeling structure and dynamics
  of a broad range of material systems across variable scales. This
  generalized phase field crystal type approach is based upon the complete
  expansion of particle direct correlation functions and the concept of
  isotropic tensors. Through applications to the modeling of various two-
  and three-dimensional crystalline structures, our study demonstrates the
  capability of bond angle control in this continuum field theory and its
  effects on the emergence of ordered phases, and provides a systematic
  way of tunable angle analysis for crystalline microstructures.
\end{abstract}

\maketitle

One of the long-lasting challenges in materials study is how to effectively
tackle the complex structural and dynamical phenomena involving multiple
spatial and temporal scales. Of particular importance is the bridging
between atomic-level microstructural details and mesoscopic, nonequilibrium
characteristics, such as mesoscale surface patterns or interface structures
that are governed by system elasticity and plasticity and by diffusional
or displacive dynamic processes. This requires novel theoretical efforts
particularly those based on coarse-graining methods beyond the traditional
single-scale atomistic or continuum approaches.
Among them much work has been devoted to the development of density-field
based schemes across different scales, as featured by the incorporation of
crystalline and microscopic attributes into probability density description
\cite{re:elder02,*re:elder04,re:elder07,Li11,re:jin06}.

Many of these field-based models can be connected to
the classical density functional theory (CDFT) \cite{re:singh91,re:lowen94}.
Through coarse-graining or ``smoothing'' the local density field
over atomic vibrational scales, the small-scale limitation of CDFT
can be mitigated, resulting in a continuum field theory with atomic
or particle-scale spatial resolution and diffusive time scales.
A fast-growing and widely applied version of such a theory is
the phase field crystal (PFC) method
\cite{re:elder02,*re:elder04,re:elder07,re:teeffelen09,re:greenwood10,wu10,
Mkhonta13,*Mkhonta16,Schwalbach13,Kocher15,re:huang10b,re:spatschek10},
with applications across a variety of solid and soft-matter
systems, particularly for the elastoplastic phenomena that are
inaccessible to traditional methods \cite{emmerich12,chan10,TothPRL11,
Ofori-opoku13,backofen14,BerryPRB15,HirvonenPRB16,Huang16,*Huang13,Taha17}.
Most PFC models are constructed for systems of isotropic interactions,
with lattice symmetry controlled by microscopic length scales
\cite{re:elder02,*re:elder04,re:elder07,re:teeffelen09,re:greenwood10,
wu10,Mkhonta13,*Mkhonta16,Schwalbach13,Kocher15,re:huang10b,re:spatschek10,
ChanPRE15,SubramanianPRL16,AlsterPRE17}. They are applicable to
metallic-type materials or colloidal systems with excluded volume or
steric interactions that are dependent on interparticle distance,
but would be a crude approximation if applied to a broader range
of material systems with directional interaction depending on both
bond lengths and angles. It is thus important to build the bond (or
particle-neighboring) angle dependency into continuum modeling which,
however, is nontrivial, given that microscopically the corresponding
interparticle interactions are \textit{anisotropic}, while
\textit{rotational invariance} of the whole system must be maintained
in the free energy functional.

In the traditional density-field approach based on Landau theory,
an additional bond-orientational order parameter and the associated
rotationally invariant orientational free energy were introduced for glassy
\cite{SteinhardtPRB83} or quasicrystalline \cite{JaricPRL85} systems.
On the other hand, in principle the orientational information should 
already be incorporated in the density functional and direct correlation
functions, although it is challenging to identify and control. The
related attempts are rather limited, and usually accompanied by
some specific assumptions as in two types of angle-dependent PFC models
developed recently. The first one \cite{wu10b} adopts some nonlinear
free-energy gradient terms introduced in previous studies of square
convection pattern \cite{Proctor81,*Gertsberg81,Bestehorn92},
while the second type is built on some pre-assumed infinite series
expansions of three-point direct correlation function $C^{(3)}$, either
through a separation of $C^{(3)}$ in real space \cite{Seymour16} or in
terms of Legendre polynomials in Fourier space \cite{AlsterPRMater17}.

Here we provide a systematic study of angular dependence and orientation
control in density-field formulation. Our analysis is based on the
property of isotropic tensor and the complete Fourier expansion of any
$n$-point direct correlation function $C^{(n)}$ that satisfies the condition
of rotational invariance, without any pre-assumptions. 
Our results show that any finite order contributions of $C^{(3)}$ expansion
to the rotationally invariant free energy are always angle independent,
as a result of the resonant condition of wave vector triads, while those
from at lease four-point correlation are needed to explicitly incorporate
the dependency on the angle between neighboring constituent particles.
Applications of this new PFC-type model include some examples of
three-dimensional (3D) structure modeling (such as simple cubic and
diamond cubic phases) via a single length scale combined with angle-dependent
effects, and importantly, the achieving of continuous angle control in
both two-dimensional (2D) and 3D crystalline structures such as 2D rhombic
and square and 3D simple monoclinic and orthorhombic phases, which
demonstrates the advantage of this angle-adjustable density field approach.

In CDFT the free energy functional is expanded via direct correlation
functions \cite{re:singh91,re:lowen94}, i.e.,
\begin{eqnarray}
  && \Delta F / k_BT = \rho_0 \int d\mathbf{r} (1+n) \ln (1+n)
  - \sum_{m} \frac{1}{m!} \rho_0^m \nonumber\\
  && \times \int \prod_{j=1}^m d\mathbf{r}_j
  ~C^{(m)}(\mathbf{r}_1, \mathbf{r}_2, ..., \mathbf{r}_m)
  n(\mathbf{r}_1) n(\mathbf{r}_2) \cdots n(\mathbf{r}_m),
\label{eq:F}
\end{eqnarray}
where $n=(\rho-\rho_0)/\rho_0$ is the density variation field, with $\rho$
the local atomic number density and $\rho_0$ a reference state density.
The condition of rotational invariance needs to be maintained for
any $m$-point direct correlation function $C^{(m)}$ and its Fourier transform
$\hat{C}^{(m)}(\mathbf{q}_1, \mathbf{q}_2, ..., \mathbf{q}_{m-1})$.
If expanding $\hat{C}^{(m)}$ as a power series of wave vector $\mathbf{q}_i$,
the resulting terms are of form $\prod_{i=1}^{m-1} \prod_{\alpha=x,y,z}
q_{i\alpha}^{n_{i\alpha}}$ ($n_{i\alpha}=0,1,2,...$), the majority of which
are, however, not rotationally invariant.
Alternatively, this expansion can be expressed in an equivalent form
$\hat{C}^{(m)}(\mathbf{q}_1, \mathbf{q}_2, ..., \mathbf{q}_{m-1})
= \sum_{K=0}^\infty \sum_{i_1... =1}^{m-1} \sum_{\alpha_{i_1}... =x,y,z}
C_{i_1\alpha_{i_1} ... i_K\alpha_{i_K}} ~ T_{i_1\alpha_{i_1} ... i_K\alpha_{i_K}}^{(K)}$,
where $T_{i_1\alpha_{i_1} ... i_K\alpha_{i_K}}^{(K)}
= q_{i_1\alpha_{i_1}} \cdots q_{i_K\alpha_{i_K}}$ can be viewed as
components of a tensor $\mathbf{T}^{(K)}$ of rank $K$. Thus the rotational
invariance condition of this expansion would be satisfied if these tensor
components are invariant under proper orthogonal group $O^+(2)$ or $O^+(3)$
transformation (i.e., 2D or 3D rotation), which is the definition of
an isotropic Cartesian tensor. Given the property of isotropic tensors
which can be written as linear combinations of products of Kronecker deltas
$\delta_{\alpha_i\alpha_j}$ (for even rank $K$) or their product with only one
Levi-Civita permutation tensor $\epsilon_{\alpha_k\alpha_l\alpha_p}$ (for odd
$K$, with $\alpha_k,\alpha_l,\alpha_p=x,y,z$) in 2D or 3D Euclidean space
\cite{JeffreysPCPS73,KearsleyJRNBSB75,ApplebyGMJ87}, the corresponding
rotationally invariant form of $\hat{C}^{(m)}$ expansion can be expressed
in terms of $\mathbf{q}_i \cdot \mathbf{q}_j$ and
$(\mathbf{q}_{k} \times \mathbf{q}_{l}) \cdot \mathbf{q}_{p}$, i.e.,
\begin{eqnarray}
  &&\hat{C}^{(m)}(\mathbf{q}_1, \mathbf{q}_2, ..., \mathbf{q}_{m-1})
  = \sum_{\mu_{11},...=0}^{\infty} \hat{C}^{(m)}_{\mu_{11}...}
  \prod_{i,j=1}^{m-1} \left (\mathbf{q}_i \cdot \mathbf{q}_j \right )^{\mu_{ij}}
  \nonumber\\
  && + \sum_{k,l,p=1}^{m-1} \sum_{\nu_{11},...=0}^{\infty} \hat{C}^{(m)}_{\nu_{11}...klp}
  \prod_{i,j=1}^{m-1} \left (\mathbf{q}_i \cdot \mathbf{q}_j \right )^{\nu_{ij}}
  \left [ \left (\mathbf{q}_{k} \times \mathbf{q}_{l} \right ) \cdot \mathbf{q}_{p}
    \right ], \nonumber\\
\label{eq:C_expan}
\end{eqnarray}
with coefficients $\hat{C}^{(m)}_{\mu_{11}...}$ and $\hat{C}^{(m)}_{\nu_{11}...klp}$.
Note that this is a general form of expansion but not an irreducible one.

For two-point correlation, from Eq.~(\ref{eq:C_expan}) with $m=2$ the only
available expansion form is $(\mathbf{q} \cdot \mathbf{q})^M = q^{2M}$, i.e.,
$\hat{C}^{(2)}(\mathbf{q}) = \hat{C}_0 + \sum_{M=1}^{\infty} \hat{C}_M q^{2M}$.
Its contribution to the free energy functional is given by (after rescaling)
\begin{equation}
  \Delta \mathcal{F}^{(2)} = \int d\mathbf{r} \left \{ -\frac{\epsilon}{2} n^2
  + \frac{\lambda}{2} n \prod_{i=0}^{N-1} \left [ \left ( \nabla^2 + Q_i^2 \right )^2
  + b_i \right ] n \right \}, \label{eq:F2}
\end{equation}
where $\epsilon$, $\lambda$, $Q_i$, and $b_i$ can be expressed via the expansion
coefficients $\hat{C}_M$.
This leads to the multi-mode PFC model presented in Ref. \onlinecite{Mkhonta13},
with wave numbers $Q_i$ determining $N$ different length scales (bond lengths).

When $m=3$, the general form of $\hat{C}^{(3)}(\mathbf{q}_1, \mathbf{q}_2)$ reads
\begin{eqnarray}
  && \hat{C}^{(3)}(\mathbf{q}_1, \mathbf{q}_2) = \hat{C}^{(3)}_0 + \sum_{M=1}^{\infty}
  \left [ \hat{C}^{(3)}_1 q_1^{2M} + \hat{C}^{(3)}_2 q_2^{2M} \right. \nonumber\\
  && + \sum_{\mu=1}^{M-1} \hat{C}^{(3)}_{2\mu, 2M-2\mu} q_1^{2\mu} q_2^{2M-2\mu}
    \label{eq:C3_expan}\\
  && \left. + \sum_{\mu=0}^{M-1}\sum_{\nu=0}^{M-1-\mu} \hat{C}^{(3)}_{2\mu, 2\nu, 2M-2\mu-2\nu}
    q_1^{2\mu} q_2^{2\nu} (\mathbf{q}_1 \cdot \mathbf{q}_2)^{M-\mu-\nu} \right ],
  \nonumber
\end{eqnarray}
with the corresponding free energy contribution given by (see Supplemental
Material (SM) \cite{SM} for the derivation)
\begin{eqnarray}
  && \Delta \mathcal{F}^{(3)} = \int d\mathbf{r} \left \{ -\frac{1}{3} D_0 n^3
  + \sum_{M=1}^{\infty} \left [ D_{M} n^2 \nabla^{2M} n \right. \right. \nonumber\\
  && + \sum_{\mu=1}^{M-1} D_{\mu,M-\mu} n (\nabla^{2\mu} n) (\nabla^{2M-2\mu} n)
     + \sum_{\mu=1}^{M-1} \sum_{\nu=1}^{M-1-\mu} \nonumber\\
  && \left. \left. \times
    D_{\mu,\nu,M-\mu-\nu} \left ( \nabla^{2\mu} n \right ) \left ( \nabla^{2\nu} n \right )
    \left ( \nabla^{2M-2\mu-2\nu} n \right ) \right ] \right \}, \label{eq:F3}
\end{eqnarray}
where parameters $D$'s are dependent on the $\hat{C}^{(3)}$ coefficients.
Interestingly, Eq.~(\ref{eq:F3}) shows that any terms of $C^{(3)}$ free
energy contribution are always angle independent and isotropic (except for some
special infinite series of $\hat{C}^{(3)}$ expansion \cite{Seymour16,AlsterPRMater17};
see the SM \cite{SM}). This can be attributed to the fact that the cubic energy
terms are governed by the resonant triads of reciprocal lattice vectors
\cite{Alexander78,Mkhonta13}, i.e., $\mathbf{q}_j + \mathbf{q}_k + \mathbf{q}_l = 0$,
and the side lengths of this vector triangle ($|\mathbf{q}_j|$, $|\mathbf{q}_k|$,
$|\mathbf{q}_l|$; i.e., lattice length scales) uniquely determine all three angles
between wave vectors and hence the bond (neighboring) orientations.

Thus we need four- or higher-order direct correlation to obtain the explicit
angle dependence, given that angles of a wave vector polygon or skew polygon of
more than 3 sides (with resonant condition $\sum_{i=1}^m \mathbf{q}_i=0$,
$m \geq 4$) cannot be uniquely determined by the side lengths. For $\hat{C}^{(4)}$
all the free energy terms are derived in the SM, including two types of isotropic
terms, $n^4$ and $n (\nabla^{2\mu} n) (\nabla^{2\nu} n) (\nabla^{2M-2\mu-2\nu} n)$
(with integers $\mu, \nu \geq 0$, $M \geq \mu+\nu$), and three types of
angle-dependent terms,
\begin{eqnarray}
  && f_{a1}^{(4)} = \left [ \nabla^{2\mu} \left ( n \nabla^{2\omega} n \right ) \right ] 
  \left (\nabla^{2\nu} n \right ) \left (\nabla^{2M-2\mu-2\nu-2\omega} n \right ), \nonumber\\
  && f_{a2}^{(4)} = \sum_{\alpha_{i},\beta_{j}=x,y,z} \left [ \nabla^{2\mu}
    \left ( n \nabla^{2\omega} \prod_{i=1}^\kappa \prod_{j=1}^\tau
    \partial_{\alpha_{i}} \partial_{\beta_{j}} n \right ) \right ] \nonumber\\
  && \times \left ( \nabla^{2\nu} \prod_{i=1}^\kappa \partial_{\alpha_{i}} n \right )
  \left ( \nabla^{2M-2\mu-2\nu-2\omega-2\kappa-2\tau} \prod_{j=1}^\tau \partial_{\beta_{j}} n
  \right ), \nonumber\\
  && f_{a3}^{(4)} = n \sum_{\alpha_{i},\beta_{j},\gamma_{k}} \sum_{\alpha,\beta,\gamma}
    \epsilon_{\alpha\beta\gamma} \left ( \nabla^{2\mu} \prod_{i=1}^\kappa \prod_{k=1}^\lambda
    \partial_{\alpha_{i}} \partial_{\gamma_{k}} \partial_{\alpha} n \right ) \nonumber\\
    && \times \left ( \nabla^{2\nu} \prod_{i=1}^\kappa \prod_{j=1}^\tau \partial_{\alpha_{i}}
    \partial_{\beta_{j}} \partial_{\beta} n \right ) 
    \left ( \nabla^{2\omega} \prod_{j=1}^\tau \prod_{k=1}^\lambda \partial_{\beta_{j}}
    \partial_{\gamma_{k}} \partial_{\gamma} n \right ), \nonumber
\end{eqnarray}
where $\alpha_{i},\beta_{j},\gamma_{k},\alpha,\beta,\gamma = x,y,z$.
For the example of $f_{a1}^{(4)}$ terms (with integers $\mu \geq 1$, $\nu, \omega
\geq 0$ and $M \geq \mu+\nu+\omega$), if expanding the density field as
$n = n_0 + \sum_j A_j \exp(i \mathbf{q}_j \cdot \mathbf{r})$, with the average
density $n_0$ and amplitudes $A_j(\mathbf{q}_j) = A_{-j}^*(-\mathbf{q}_j)$,
given a system of volume $V$ we have
\begin{eqnarray}
  && \frac{1}{V} \left. \int d\mathbf{r} f_{a1}^{(4)} \right |_{n_0=0}
  = (-1)^M \sum_{ijkl} |\mathbf{q}_i+\mathbf{q}_j|^{2\mu} \nonumber\\
  && \quad \times q_j^{2\omega} q_k^{2\nu} q_l^{2M-2\mu-2\nu-2\omega} A_i A_j A_k A_l
  ~\delta_{\mathbf{q}_i+\mathbf{q}_j+\mathbf{q}_k+\mathbf{q}_l,0}. \label{eq:f_C4}
\end{eqnarray}
The resonant condition $\mathbf{q}_i+\mathbf{q}_j+\mathbf{q}_k+\mathbf{q}_l=0$
is satisfied by 3 types of wave vector combinations \cite{Jones94,Mkhonta16}:
collinear ($\mathbf{q}_i-\mathbf{q}_i+\mathbf{q}_i-\mathbf{q}_i=0$), pairwise
($\mathbf{q}_i-\mathbf{q}_i+\mathbf{q}_j-\mathbf{q}_j=0$), and nonpairwise closed
loops. For Eq.~(\ref{eq:f_C4}), the collinear contribution $f^C$ from $n_q$ wave
vectors yields $f^C=(-1)^M 2^{2\mu+1} \sum_{j=1}^{n_q} q_j^{2M} |A_j|^4$, while the
angle dependence arises from the factor $|\mathbf{q}_i+\mathbf{q}_j|^{2\mu}$ if
$\mu \ge 2$ for pairwise resonant tetrads and $\mu \ge 1$ for nonpairwise ones.

For some crystalline structures (e.g., five 2D Bravais lattices and some
3D ones) the pairwise contributions would be sufficient in determining the phase
stability. Given any pair $(\mathbf{q}_i, \mathbf{q}_j)$ with angle $\theta$
and $q_j = \gamma q_i \equiv \gamma q$ ($i \neq j$), the pairwise (P) contribution
of Eq.~(\ref{eq:f_C4}) gives
\begin{eqnarray}
  && \frac{1}{V} \left. \int d\mathbf{r} \left [ \nabla^{2\mu} (n \nabla^{2\omega} n)
    \right ] (\nabla^{2\nu} n) (\nabla^{2M-2\mu-2\nu-2\omega} n) \right |_{n_0=0}^{(P)}
  \nonumber\\
  && = 2 (-1)^M \left [ (1+\gamma^2 + 2\gamma\cos\theta)^\mu
    + (1+\gamma^2 - 2\gamma\cos\theta)^\mu \right ]  \nonumber\\
  && \times q^{2M} (\gamma^{2M-2\mu-2\nu-2\omega}+\gamma^{2\nu}) (1+\gamma^{2\omega})
  |A_i|^2 |A_j|^2, \label{eq:f_C4_P}
\end{eqnarray}
the minimization of which leads to $\cos \theta = 0$ when $(-1)^M > 0$. A similar
outcome is obtained for all other angle-dependent $\hat{C}^{(4)}$ contributions
\cite{SM}, indicating that a single quartic gradient term would favor only the
$\pi/2$ orientation when considering pairwise wave vectors.

\begin{figure}
  \centerline{\includegraphics[width=0.5\textwidth]{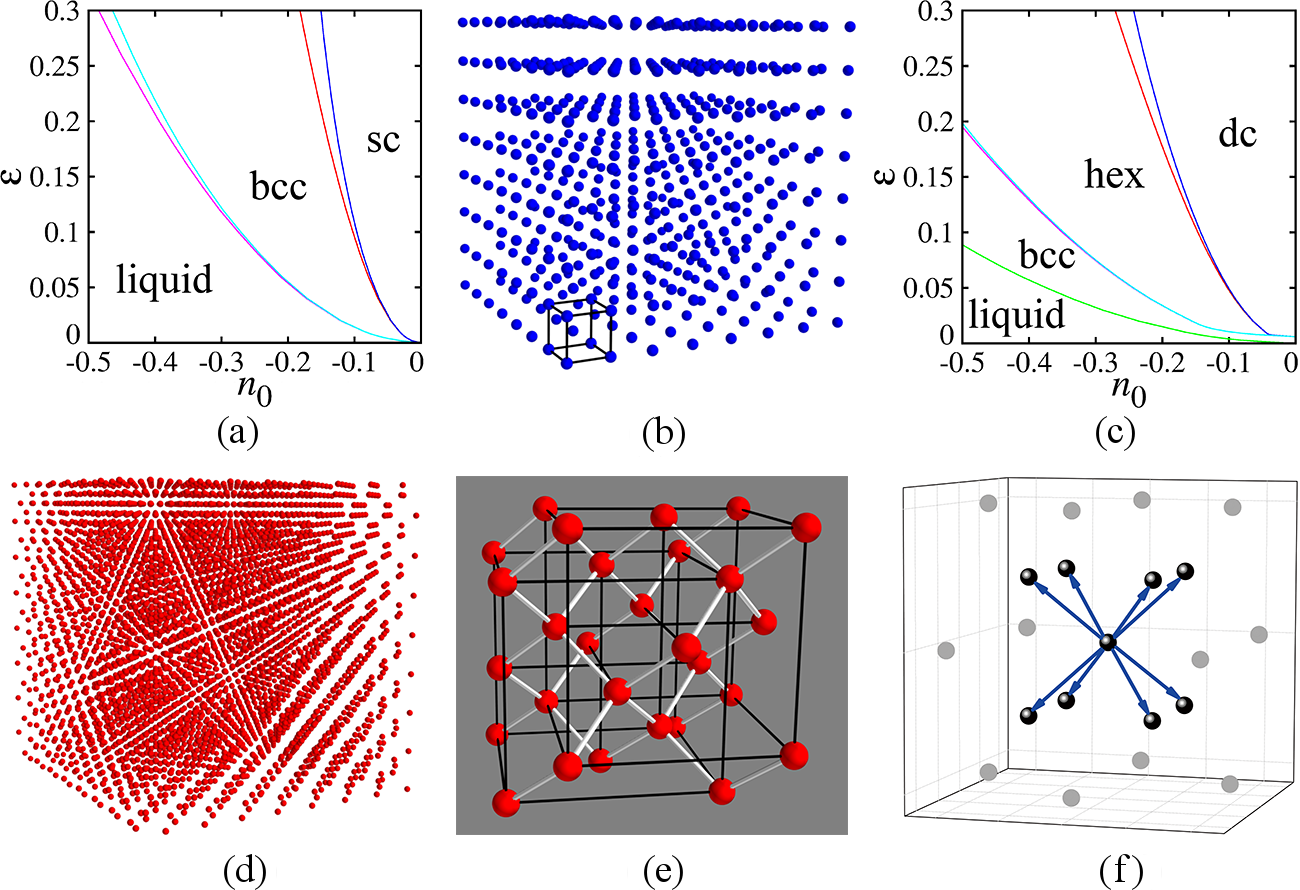}}
  \caption{Phase diagrams determined by Eq. (\ref{eq:F_2_4}) at
    $\lambda=1$ and $(E_0, E_{11}, E_{44}) = (1, 25/72, 1/16)$ for (a)
    and $(1/18, 0, 1/32)$ for (c).
    Sample sc (b) and dc (d) structures are obtained from
    simulations with $64^3$ grid size, for $n_0=-0.01$ and $\epsilon=0.02$.
    An enlarged portion of (d) is shown in (e), while (f) gives the diffraction
    pattern of (d).}
\label{fig:sc_dc}
\end{figure}

To verify this result we study a 3D example with only one length scale (i.e.,
one mode with $q_0=1$ and $\gamma=1$),
\begin{eqnarray}
  \mathcal{F} = \int d\mathbf{r} && \left [ -\frac{1}{2} \epsilon n^2
    + \frac{1}{2} \lambda n \left ( \nabla^2 + q_0^2 \right )^2  n
    + \frac{1}{4} E_0 n^4 \right. \nonumber\\
  && \left. + E_{11} n^3 \nabla^2 n + E_{44} n^2 \nabla^4 n^2 \right ], \label{eq:F_2_4}
\end{eqnarray}
where the only angle-dependent term is $n^2 \nabla^4 n^2$ (i.e., $M=\mu=2$
and $\nu=\omega=0$, reproducing that used previously in 2D square pattern
formation \cite{Proctor81,*Gertsberg81,Bestehorn92}).
From the above analysis the simple cubic (sc) phase, characterized by a bond
angle $\theta=\pi/2$ and basic wave vectors $(1,0,0)$, $(0,1,0)$, $(0,0,1)$,
should be stabilized for $n_0$ close to 0. This is consistent with our numerical
result in Fig.~\ref{fig:sc_dc}(b), for which the simulation starts from a
homogeneous state with random initial condition and follows the dynamics 
$\partial n / \partial t = \nabla^2 \delta \mathcal{F} / \delta n$.
The phase diagram is given in Fig.~\ref{fig:sc_dc}(a), as calculated via
one-mode approximation.

To model structures characterized by other angles, the parameters need
to be chosen such that the contributions from nonpairwise wave vectors would
be important. For the example of diamond cubic (dc) phase, in the first mode
with amplitude $A$, $\mathbf{q}_1=q_0(-1, 1, 1)/\sqrt{3}$,
$\mathbf{q}_2=q_0(1, -1, 1)/\sqrt{3}$, $\mathbf{q}_3=q_0(1, 1, -1)/\sqrt{3}$,
and $\mathbf{q}_4=q_0(-1, -1, -1)/\sqrt{3}$; thus
$\mathbf{q}_1+\mathbf{q}_2+\mathbf{q}_3+\mathbf{q}_4=0$, yielding
$\cos \theta = -1/3$ and $\theta=109.47^\circ$. The corresponding
nonpairwise contribution of Eq.~(\ref{eq:f_C4}) is then given by
$(-1)^{M+1} 48 q_0^{2M} 2^\mu (1+\cos \theta)^\mu |A|^4$.
Combining with Eq.~(\ref{eq:f_C4_P}), we can identify the parameters minimizing
$\mathcal{F}$ of Eq.~(\ref{eq:F_2_4}) that favor the dc structure, with results
(including the phase diagram and a simulated structure emerging from initial
homogeneous state) shown in Figs.~\ref{fig:sc_dc}(c)-\ref{fig:sc_dc}(f).
Note that due to the incorporation of angle dependence, only one mode is
needed to generate sc or dc phase, different from previous isotropic PFC models
where three \cite{re:greenwood10} or two \cite{ChanPRE15} modes are required.

An important feature of this approach is the ability to
continuously control the characteristic angles of crystalline phases,
as achieved by combining angle-dependent gradient terms,
e.g., $\sum_k E_k n (\nabla^{2\mu_k} n^2) (\nabla^{2M_k-2\mu_k} n)$, so that
the angle can be tuned via coefficients $E_k$. For the case of a single
adjustable angle $\theta$ between any pair of wave vectors
$(\mathbf{q}_i, \mathbf{q}_j)$, the simplest combination is
$E_1 n (\nabla^{2\mu_1} n^2) (\nabla^{2M_1-2\mu_1} n) + E_2 n (\nabla^{2\mu_2} n^2)
(\nabla^{2M_2-2\mu_2} n)$. For the structures dominated by pairwise and collinear
wave vector contributions, Eq.~(\ref{eq:f_C4_P}) gives
\begin{eqnarray}
  && f^P = \frac{1}{V} \int d\mathbf{r} \left. \sum_{k=1}^2 E_k n (\nabla^{2\mu_k} n^2)
  (\nabla^{2M_k-2\mu_k} n) \right |_{n_0=0}^{(P)} \nonumber\\
  && = 4 \left \{ E_1 \left [ (1+\gamma^2 + 2\gamma\cos\theta)^{\mu_1}
    + (1+\gamma^2 - 2\gamma\cos\theta)^{\mu_1} \right ] \right. \nonumber\\
  && \qquad \times (-1)^{M_1} q^{2M_1} (1+\gamma^{2M_1-2\mu_1}) \nonumber\\
  && \quad + E_2 \left [ (1+\gamma^2 + 2\gamma\cos\theta)^{\mu_2}
    + (1+\gamma^2 - 2\gamma\cos\theta)^{\mu_2} \right ] \nonumber\\
  && \qquad \left. \times (-1)^{M_2} q^{2M_2} (1+\gamma^{2M_2-2\mu_2}) \right \}
  |A_i|^2 |A_j|^2, 
\end{eqnarray}
while the collinear contribution is angle independent, i.e.,
$f^C = \sum_{k=1}^2 (-1)^{M_k} q^{2M_k} 2^{2\mu_k+1} E_k ( |A_i|^4 + \gamma^{2M_k}|A_j|^4 )$.
By minimizing $f^P$ we get $\sin \theta = 0$ or
\begin{eqnarray}
  && -\frac{E_2'}{E_1'} \left [ (1+\gamma^2 + 2\gamma\cos\theta)^{\mu_2-1}
    - (1+\gamma^2 - 2\gamma\cos\theta)^{\mu_2-1} \right ] \nonumber\\
  && = (1+\gamma^2 + 2\gamma\cos\theta)^{\mu_1-1} -
  (1+\gamma^2 - 2\gamma\cos\theta)^{\mu_1-1}, \label{eq:E2E1_theta}
\end{eqnarray}
where $E_k' = (-1)^{M_k} q^{2M_k} \mu_k E_k (1+\gamma^{2M_k-2\mu_k})$ ($k=1,2$).
It is straightforward to show that the lowest order terms giving adjustable
values of nonzero $\theta$ for $f^P$ minimization are of $\mu_1=4$ and
$\mu_2=2$, when $(-1)^{M_1} E_1 >0$ and $(-1)^{M_2} E_2 <0$;
thus $\cos^2 \theta = - [E_2'/E_1' + 3(1+\gamma^2)^2]/4\gamma^2$ from
Eq.~(\ref{eq:E2E1_theta}). To ensure the results are independent of wave number
$q$, we set $M_1=M_2=M$ and to lowest order $M=\mu_1=4$, $\mu_2=2$, leading to
the combination $E_1 n^2 \nabla^8 n^2 + E_2 n (\nabla^4 n^2) (\nabla^4 n)$ and
\begin{equation}
  \frac{E_2}{E_1} = - \frac{4}{1+\gamma^4} \left [ 3(1+\gamma^2)^2
    + 4\gamma^2 \cos^2 \theta \right ], ~ E_1>0, \label{eq:M4mu42}
\end{equation}
i.e., at least 8th-order gradient terms are needed to obtain the angle control
in structures governed by pairwise resonant wave vectors. The free energy
functional is then
\begin{eqnarray}
  && \mathcal{F} = \int d\mathbf{r} \left [ -\frac{\epsilon}{2} n^2
    + \frac{\lambda}{2} n \prod_{i=0}^{N-1} \left ( \nabla^2 + Q_i^2 \right )^2 n
    + E_1 n^2 \nabla^8 n^2 \right. \nonumber\\
  && \left. + E_2 n  ( \nabla^4 n^2  )
    (\nabla^4 n) + E_3  ( \nabla^2 n^2 ) (\nabla^4 n) (\nabla^2 n)
    + \frac{E_0}{4} n^4 \right ], \nonumber\\
  \label{eq:F_E}
\end{eqnarray}
where the $E_3$ term is angle independent for pairwise wave vectors and is
introduced for structure stability. 

\begin{figure}
\centerline{\includegraphics[width=0.5\textwidth]{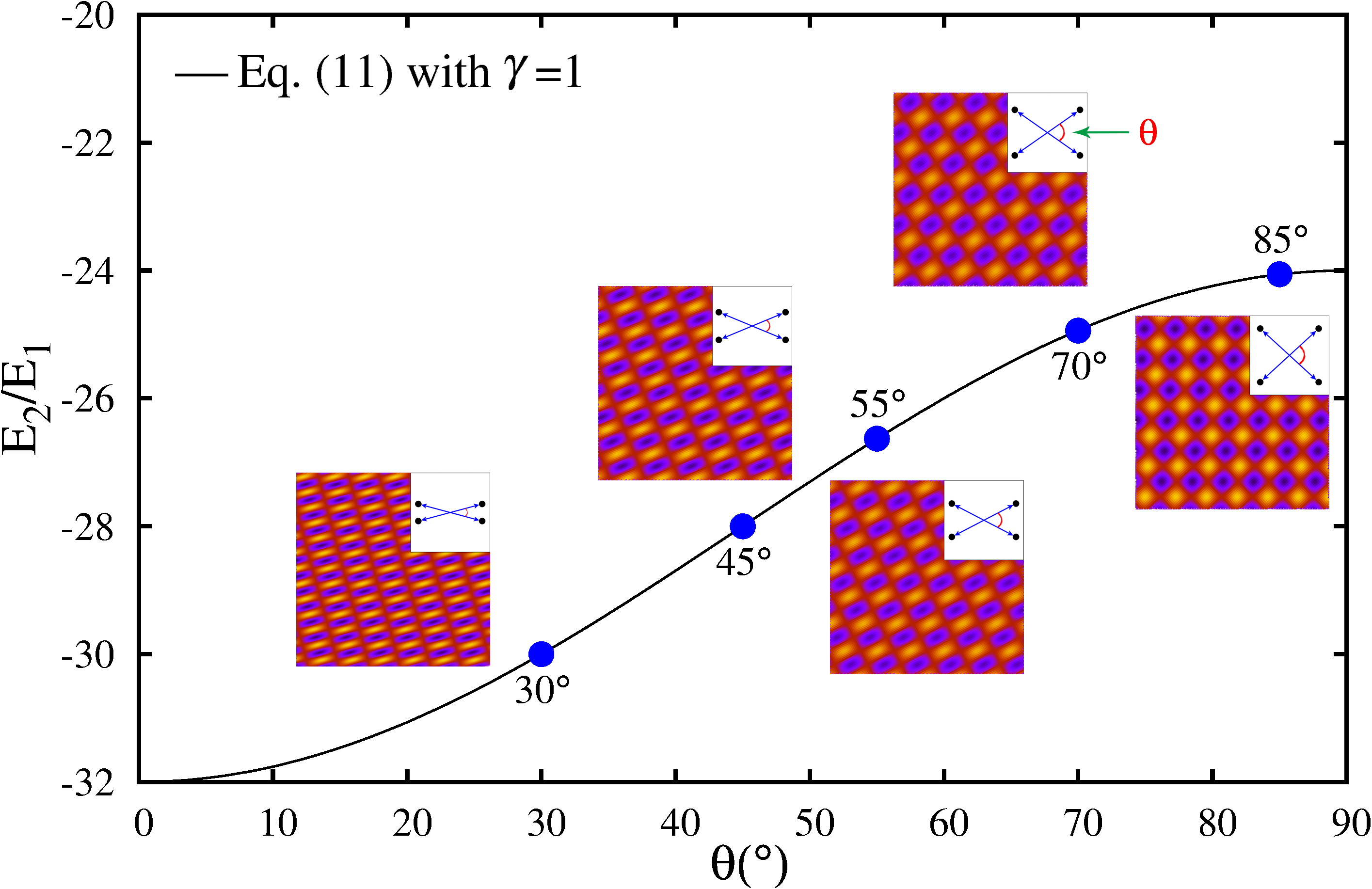}}
\caption{Angle control for rhombic phase, based on the prediction of Eq.~(\ref{eq:M4mu42})
  for $E_2/E_1$ vs $\theta$ (solid curve). Simulated structures and diffraction patterns
  are obtained with $n_0=0$, $\epsilon=0.01$, $E_0=1/3$, $E_3=0$, and $(\lambda, E_1)
  = (600, 1/750)$ for $\theta=85^{\circ}$ and $70^{\circ}$, $(2 \times 10^4, 1/750)$ for
  $\theta=55^{\circ}$, $(6 \times 10^4, 1/800)$ for $\theta=45^{\circ}$, and
  $(6 \times 10^5, 1.104 \times 10^{-3})$ for $\theta=30^{\circ}$.}
\label{fig:rhombic}
\end{figure}

We first apply the above analysis to the modeling of 2D rhombic and square
phases with continuous angle selection ($0 < \theta \leq \pi/2$), using
one mode with $N=Q_0=\gamma=1$ and the basic wave vectors
$\mathbf{q}_{1,2} = (\mp \cos(\theta/2), \sin(\theta/2))$. 
Some simulation results are illustrated in Fig.~\ref{fig:rhombic},
for 5 sample rhombic structures with $\theta = 30^{\circ}, 45^{\circ},
55^{\circ}, 70^{\circ}, 85^{\circ}$, starting from homogeneous initial state.
The parameter ratio $E_2/E_1$ is chosen according to Eq.~(\ref{eq:M4mu42}),
and $E_3=0$. The resulting structures with desired angles are corroborated by
the associated diffraction patterns (Fig.~\ref{fig:rhombic} insets), indicating
the capability of angle control via nonlinear gradient terms. 

\begin{figure}
\centerline{\includegraphics[width=0.5\textwidth]{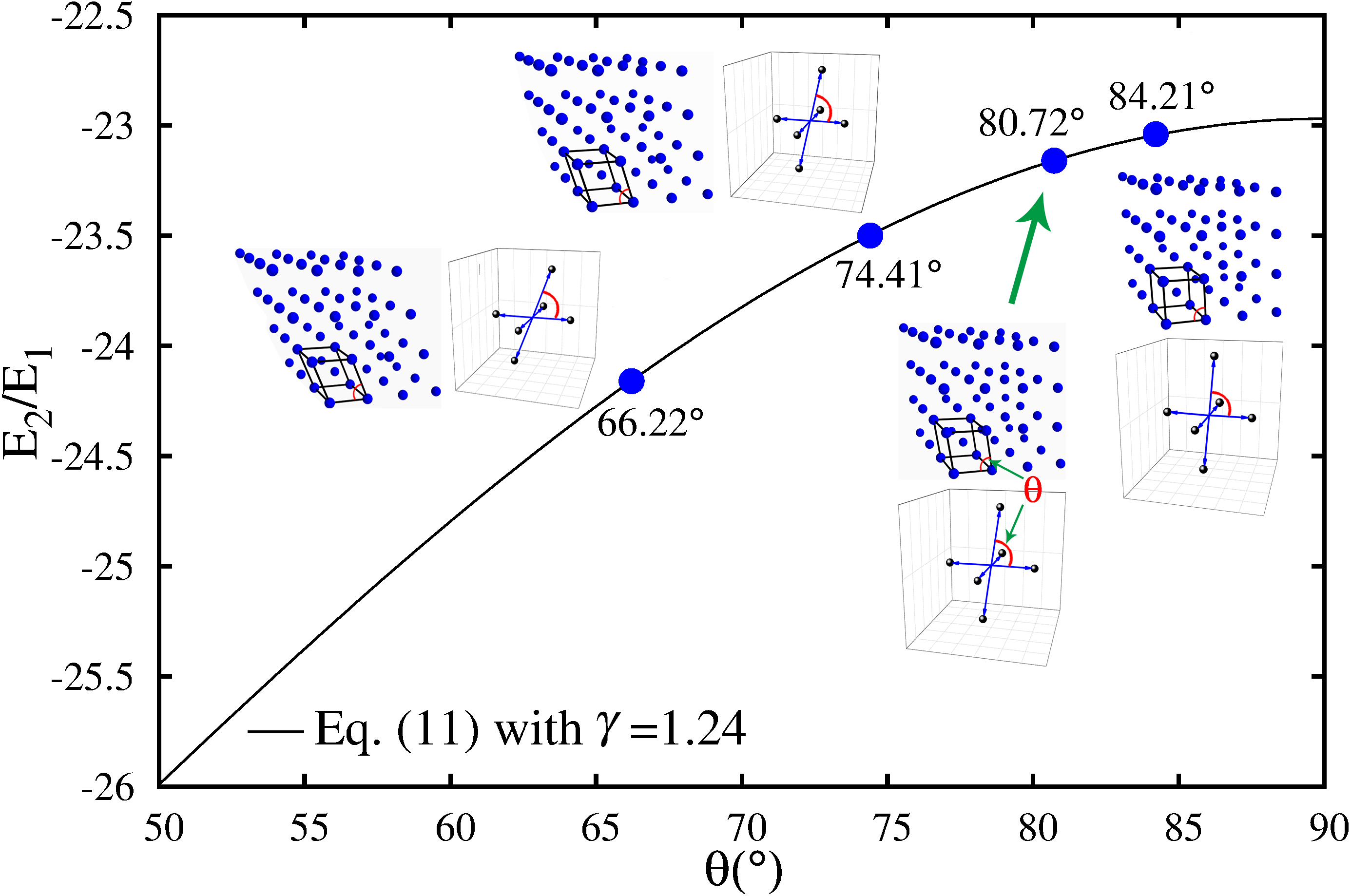}}
\caption{Angle control for simple monoclinic structures with $Q_0:Q_1:Q_2
  =1:1.16:1.24$, based on the prediction of Eq.~(\ref{eq:M4mu42}). A portion
  of simulated system and the diffraction pattern are shown for each angle,
  with $n_0=0$, $\epsilon=0.01$, $E_0=1/2$, $E_1=1/240$, $E_3=5/32$, and
  $\lambda=5 \times 10^6$ for $\theta = 66.22^{\circ}$ and $5 \times 10^5$
  for $\theta = 74.41^{\circ}$, $80.72^{\circ}$, $84.21^{\circ}$.}
\label{fig:monoclinic}
\end{figure}

Similar outcomes of continuous angle control can be obtained in 3D from
Eqs.~(\ref{eq:M4mu42}) and (\ref{eq:F_E}), with an example of simple monoclinic
phase presented in Fig.~\ref{fig:monoclinic}. Three modes, $Q_0:Q_1:Q_2
=1:\gamma_2:\gamma_3$, are needed here, with basic wave vectors
$\mathbf{q}_1=(1,0,0)$, $\mathbf{q}_2=(0,\gamma_2,0)$, and $\mathbf{q}_3=
(\gamma \cos \theta,0,\gamma \sin \theta)$ with $\gamma \equiv \gamma_3$.
This gives $\theta_{12} = \theta_{23} = \pi/2$, where $\theta_{ij}$ is the angle
between $\mathbf{q}_i$ and $\mathbf{q}_j$, and $\theta_{13} \equiv \theta$ is
the only tunable angle determined by Eq.~(\ref{eq:M4mu42}). The corresponding
structures of different $\theta$, including simple orthorhombic with
$\theta=\pi/2$, have been obtained in our numerical simulations using random
initial and periodic boundary conditions (Fig.~\ref{fig:monoclinic}).
Note that this modeling procedure is also applicable to
other angle-adjustable phases, such as
rhombohedral (trigonal) 
or more complex case of triclinic (with 3 modes and 3 tunable angles). All these
results thus verify the effect of angle tuning and control on the emergence of
crystalline phases through contributions of 
quartic coupling.

It is also important to note that although the model introduced above involves
high-order nonlinear gradient terms, the related computational cost is modest
when using the pseudospectral numerical algorithm, particularly for the cases
of weak segregation (i.e., small $\epsilon$) simulated here. In addition, 
such a format with spatial gradient terms has the advantage of being more
feasible for the construction of amplitude equation formalism describing slowly
varying mesoscopic scales \cite{re:huang10b,re:spatschek10,Ofori-opoku13,Huang16,*Huang13},
which is important for large-scale simulations with high computational efficiency
and is the subject of our future research.

In summary, we have constructed a complete density field formulation integrating
the microscopic property of interparticle bond-angle anisotropy and the requirement
of global-scale system rotational invariance. Our results demonstrate that effects
of angle dependency and adjustment are incorporated explicitly through quartic
correlation in the system, but not through any finite-order cubic
coupling which instead implicitly affects angle selection via lattice length
scales. The resulting nonlinear gradient terms of atomic density field have been
utilized to model various crystalline phases and importantly, their bond angle
control. Since the model developed here already incorporates system elasticity
and plasticity as in other PFC-type models, it can be readily applied to the
study or prediction of a broad range of crystalline or polycrystalline material
systems and more complex phases with bond anisotropy,
their elastoplastic and defect properties, and nonequilibrium phenomena
during crystallization and growth. This approach is built on the full-order
expansion of direct correlation functions and the application of isotropic
Cartesian tensor, and is thus of generic nature and applicable to different
types of ordering or self-assembling systems with varying atomistic details.

\begin{acknowledgments}
  Z.-F.H. acknowledges support from the National Science Foundation under Grant
  No. DMR-1609625. Z.R.L. was supported by the National Natural Science Foundation
  of China under Grant No. 21773002.
\end{acknowledgments}

\bibliography{angle_pfc_references}

\end{document}